%

\def\lsim{\mathrel{\scriptstyle{\buildrel < \over \sim}}}
\magnification 1200
\baselineskip=17pt


\centerline{\bf SPIN-DENSITY WAVE IN ISING-COUPLED ANTIFERROMAGNETIC CHAINS}
\bigskip
\vskip 50pt
\centerline{J. P. Rodriguez,$^{(a)}$ P.D. Sacramento,$^{(b,c)}$
and V.R. Vieira$^{(c)}$}
\medskip
\centerline{$^a${\it Instituto de Ciencia de Materiales,
Consejo Superior de Investigaciones Cientificas,}}
\centerline{{\it Universidad Autonoma de Madrid,
Cantoblanco, 28049 Madrid, Spain} 
{\it and}}
\centerline{\it Dept. of Physics and Astronomy,
California State University,
Los Angeles, CA 90032, USA.}
\smallskip
\centerline{$^b${\it Departamento de F\'{\i}sica, Instituto Superior T\'ecnico,
1096 Lisboa Codex, Portugal.}}
\smallskip
\centerline{$^c${\it Centro de F\'{\i}sica das Interac\c c\~ oes Fundamentais,
Instituto Superior T\'ecnico,}}
\centerline{\it 1096 Lisboa Codex, Portugal.}
\vskip 30pt
\centerline  {\bf  Abstract}
\vskip 8pt\noindent
The effect of anisotropy in the nearest-neighbor spin interactions
that couple $N\geq 2$ consecutive spin-1/2 antiferromagnetic chains is studied
theoretically by considering the limit where the coupling
is purely of the Ising
type.  An analysis based on the equivalent Luttinger model reveals
that the groundstate is an Ising antiferromagnet
in general.
\bigskip
\noindent
PACS Indices: 75.10.Jm, 75.30.Fv, 75.30.Gw, 74.20.Mn
\vfill\eject

The recent discovery of the ``ladder'' materials spawned from research in
high-temperature superconductivity has renewed interest in the physics of
coupled antiferromagnetic chains.$^1$  The former systems are composed of
magnetically isolated ladders of spin-1/2 moments that experience an
effective exchange interaction between all nearest neighbors.  Both
experiment and theory find that the $S=1/2$ antiferromagnetic ladder
shows a spin gap of order the exchange coupling constant between 
chains.$^{1,2}$  
This result, however, should be compared with that corresponding
to a single chain, which exhibits no spin gap.$^3$  In general,
an analysis of the weak
and of the strong-coupling limits reveals 
that no spin gap appears for an  odd number of chains.
	
Deviations from isotropy in the Heisenberg spin coupling that results
from the exchange interaction can occur naturally, however, and have
interesting theoretical consequences.  It is known, for example,
that Ising anisotropy produces a spin gap in the case of a single
chain.$^{4-6}$  In the present context of coupled antiferromagnetic 
chains, the following question then arises:  What effect does
anisotropy in the Heisenberg spin couplings that run 
{\it perpendicular} to the spin-1/2 chains  have on the ground state?
In response to this query, we shall study here the low-temperature
physics of a finite number $N$ of consecutive spin-1/2 $XXZ$ 
chains$^3$ coupled via a nearest-neighbor Ising interaction. 
The calculational strategy will be     to first transcribe the
problem to that of interacting spinless fermions utilizing the
Jordan-Wigner transformation, and to then apply the well-known
abelian bosonization technique.$^{6,7}$  On this basis, we arrive
at the following conclusions valid for $N\ge 2$ Ising-coupled
antiferromagnetic chains:
({\it i})  The groundstate is in a pinned spin-density wave (SDW) 
state that is commensurate with the lattice,$^8$ and that
exhibits a spin gap; ({\it ii}) The SDW slides along the direction
parallel to the chains in the case where it 
is incommensurate,$^{9}$
which can result from the application of an external magnetic field.$^{10}$
It is noteworthy that the bosonization analysis employed here
is closely related to a
fermion analogy  for the Lawrence-Doniach model
description of layered superconductors.$^{11,12}$
We now pass to the calculations.

The Hamiltonian for $N$ Ising-coupled antiferromagnetic chains 
that are sequentially ordered  can be
divided into parallel and perpendicular parts,
$H = H_{\parallel} + H_{\perp}$, where
$$H_{\parallel} = \sum_{l = 1}^N \sum_i
[J_{xy}^{\parallel}(S_{i,l}^x S_{i+1,l}^x + S_{i,l}^y S_{i+1,l}^y) +
 J_z^{\parallel} S_{i,l}^z S_{i+1,l}^z 
+ h_z S_{i,l}^z] \eqno (1)$$
and
$$H_{\perp} = \sum_{l = 1}^{N-1} \sum_i 
J_z^{\perp} S_{i,l}^z S_{i,l+1}^z \eqno (2)$$
describe  respectively 
the spin couplings within and in between the corresponding $XXZ$
chains.  Here, 
the spin operator $\vec S_{i,l}$ acts on 
spin-1/2 states lying at the $i^{\rm th}$ site of
chain $l$, and $h_z$ denotes an external magnetic
field directed along the $z$ direction.  
We presume an antiferromagnetic sign, 
$J_{xy}^{\parallel} > 0$, for the intra-chain $XY$ coupling,
and we shall set $\hbar = 1$
throughout.  The Jordan-Wigner transformation$^{3,6}$ then yields
a system of ideal spinless fermions, with
a degenerate energy spectrum
$\varepsilon_k = - J_{xy}^{\parallel} {\rm cos}\, ka$
as a function of
momentum $k$ along each chain $l$, 
that interact through both the intra-chain and the inter-chain Ising couplings,
$J_z^{\parallel}$ and $J_z^{\perp}$.  
Here, $a$ denotes the parallel lattice constant.
In this language, an antiferromagnetic
chain  then  corresponds to a half-filled band, $\varepsilon_k < 0$.
Note that inter-chain $XY$ coupling introduces unwieldy ``string''
contributions$^{6}$ into the Hamiltonian, which is
one reason why 
it has been omitted.
The continuum limit can  be taken
{\it \`a la} Kogut and Susskind
under these conditions,$^7$  and
we thereby obtain the following
Luttinger model for the parallel and perpendicular
pieces of the  Ising-coupled antiferromagnetic chains:
$$
H_{\parallel} =  \sum_{l = 1}^N\int dx\Biggl[
J_{xy}^{\parallel} a \Bigl(L_l^{\dag}  i\partial_x L_l
- R_l^{\dag} i\partial_x R_l\Bigr)
 + 4J_z^{\parallel} a L_l^{\dag}R_l^{\dag} R_l L_l +
h_z  (L_l^{\dag} L_l +  R_l^{\dag} R_l) \Biggr]\eqno (3)$$
and $H_{\perp} = H_{\perp ,1} + H_{\perp ,3} + H_{\perp ,4}$, 
with a backscattering term
$$\eqalignno{
H_{\perp ,1} = & \sum_{l=1}^{N-1}\int dx J_z^{\perp}  a
\Bigl[
L_l^{\dag} R_{l+1}^{\dag} L_{l+1} R_l
+ {\rm H.c.}\Bigr], & (4)\cr}$$
an inter-chain   umklapp term
$$\eqalignno{
H_{\perp ,3} = & \sum_{l=1}^{N-1}\int dx J_z^{\perp} a
\Bigl[
R_l^{\dag} R_{l+1}^{\dag} L_{l+1} L_l e^{i 4 k_F x}
+ {\rm H.c.}\Bigr], & (5)\cr}$$
and  an inter-chain   forward scattering term 
$$\eqalignno{
H_{\perp ,4} = & \sum_{l=1}^{N-1}\int dx
J_z^{\perp} a
:(L_l^{\dag} L_l +  R_l^{\dag} R_l):
:(L_{l+1}^{\dag} L_{l+1} +  R_{l+1}^{\dag} R_{l+1}):. & (6)\cr}$$
Here      $e^{-ik_F x} R_l (x)$ and $e^{ik_F x} L_l (x)$ denote
field operators for right and left
moving spinless fermions that move along the $l^{\rm th}$
chain, with a Fermi surface at $\pm k_F$, while the symbols `: :' represent 
normal ordering.$^{6,7}$ 
The intra-chain umklapp term$^{4,5}$ that has been omitted 
from Eq. (3) will
be discussed at the end of the paper [see Eq. (15)].

We first consider two chains,$^{13}$ in which case the above Luttinger
model can be treated exactly.  This is achieved by observing that
the chain index can be interpreted as a pseudo spin label.
Equations (3)-(6) then describe the Luther-Emery model for
pseudo spin-1/2 fermions in this
instance.$^{14-16}$  Since such fermions experience pseudo 
spin-charge separation, we have that the coupled chains
factorize following $H_{\parallel} + H_{\perp} = H_{\rho} + H_{\sigma}$,
where 
$$\eqalignno{
H_{\rho} = & 2\pi v_{\rho} \sum_{q > 0}\sum_{j = R,L} 
\rho_j(q)\rho_j(-q) + g_{\rho}\sum_q\rho_R(q)\rho_L(-q)
+ H_{\perp,3} & (7)\cr
H_{\sigma} = & 2\pi v_{\sigma} \sum_{q > 0}\sum_{j = R,L} 
\sigma_j(q)\sigma_j(-q) + g_{\sigma}\sum_q\sigma_R(q)\sigma_L(-q)
+ H_{\perp,1} & (8)\cr}$$
are the respective commuting portions of the Hamiltonian.
Here, $\rho_j (q) = 2^{-1/2} [\rho_j(q,1) + \rho_j(q,2)]$ 
and  
$\sigma_j (q) = 2^{-1/2} [\rho_j(q,2) - \rho_j(q,1)]$ 
are the particle-hole operators for pseudo-charge and 
pseudo-spin excitations, with 
$\rho_R (q,l) = \sum_k a_l^{\dag}(q+k) a_l(k)$ and
$\rho_L (q,l) = \sum_k b_l^{\dag}(q+k) b_l(k)$. 
The operators $a_l(k)$ and $b_l(k)$ respectively annihilate right
and left moving electrons of momentum $k$ on the $l^{\rm th}$ chain.  Also,
the Fermi velocities and interaction strengths for each
component
are renormalized by the inter-chain forward scattering process (6)  to 
$$\eqalignno{
v_{\rho,\sigma} = & J_{xy}^{\parallel} \pm J_z^{\perp}/2\pi, & (9)\cr
g_{\rho,\sigma} = & 4J_z^{\parallel} \pm J_z^{\perp}, & (10)\cr}$$
where the $+(-)$ signs above correspond to the $\rho (\sigma)$ label.
Application of the bosonic representation$^7$
$$\eqalignno{
R_l (x) \cong  & (2\pi\alpha)^{-1/2} {\rm exp}[i(4\pi)^{1/2} \phi_R(x,l)] 
& (11)\cr
L_l(x)  \cong  & (2\pi\alpha)^{-1/2} {\rm exp}[-i(4\pi)^{1/2} \phi_L(x,l)] 
& (12)\cr
}$$
for the spinless fermions, where
$\phi_{j}(x, l) = {\rm lim}_{\alpha\rightarrow 0}
2\pi L_x^{-1} \sum_q q^{-1} {\rm exp}
(-{1\over 2} \alpha |q| -iqx) \rho_{j}(q,l)$ are the bosonic fields
corresponding to right and left moving fermions ($j=R,L$),
reveals that the spectrum of the
pseudo-spin sector (8) has a gap, $\Delta_{\sigma}\neq 0$,
for $-g_{\sigma} < |J_z^{\perp}|$.$^{16}$ 
(In general, $\alpha^{-1}\sim a^{-1}$ is the momentum cutoff of the
Luttinger model.)
At half-filling ($h_z = 0$), where
$4k_F = 2\pi/a$ and the umklapp process (5) is at work, similar
considerations indicate that a gap, $\Delta_{\rho}\neq 0$, opens in
the pseudo-charge spectrum (7) for $-g_{\rho} < |J_z^{\perp}|$.  Comparison
of Eq. (10) with these conditions yields the phase diagram shown
in Fig. 1 for two Ising-coupled spin-1/2 $XXZ$ chains.
We remind the reader that the spinless
fermions corresponding to the pseudo-charge 
(7) and the pseudo-spin
(8) systems  are noninteracting along the respective Luther-Emery lines
$g_{\rho} = 6\pi v_{\rho}/5$ and $g_{\sigma} = 6\pi v_{\sigma}/5$,
at which point the gaps have value 
$\Delta_{\rho,\sigma} = (a/\alpha)(|J_z^{\perp}|/2\pi)$.  On the other hand,
the Coulomb gas analogy$^{16}$ as well as mean-field theory$^{12}$
indicate that the latter vanish exponentially as the
Luttinger-liquid state is approached; e.g., 
$\Delta_{\rho,\sigma} \sim |J_{z}^{\perp}| {\rm exp}
[-2\pi v_{\rho,\sigma}/(|J_z^{\perp}| + g_{\rho,\sigma})]$.

We now address the issue of the physical character of the two-chain system
in the case where all spin coupling are antiferromagnetic,
which means that $\Delta_{\rho,\sigma}\neq 0$
(see Fig. 1).  
Since the longitudinal paramagnetic susceptibility is simply given by
the pseudo-charge compressibility, we have that
this quantity follows  the activated
behavior $\chi_z\propto {\rm exp}(-2\Delta_{\rho}/k_B T)$ 
at low temperatures.  Also, pseudo charge-spin separation
implies that the specific heat is given by the sum
$C_v = C_{\rho} + C_{\sigma}$ of the respective charge and
spin contributions, each of which follow the
activated behavior 
$C_{\rho,\sigma}\propto  {\rm exp}(-2\Delta_{\rho,\sigma}/k_B T)$
at low temperatures.  The question of long-range order at zero
temperature can be attacked with the bosonization method (11) and (12).
Following Luther and Peschel,$^3$ the 
static transverse spin
correlator on the same chain may be calculated using the formula
$$\langle S_{x,1}^+ S_{0,1}^-\rangle = e^{i 2k_F x}
\langle e^{i\pi^{1/2} [\phi_R (x,1)-\phi_L (x,1)]}
e^{-i\pi^{1/2} [\phi_R (0,1)-\phi_L (0,1)]}\rangle / 4\pi\alpha  \eqno (13)$$
that is
valid to lowest order in $e^{i 2k_F x}$.  Pseudo spin-charge separation
then implies that this correlator has the 
asymptotic form
$\langle S_{x,1}^+ S_{0,1}^-\rangle = e^{i 2k_F x} G^{(\rho)}_{TS}(x)
G^{(\sigma)}_{TS}(x)/4\pi\alpha$, where 
$G^{(\sigma)}_{TS}(x) =  (\alpha/x)^{\theta_{\sigma}} e^{-x/\xi_{\sigma}}$
is the auto-correlation
function for pseudo-triplet superconductivity,$^{15,16}$ 
with Cooper pairs of effective {\it unit} charge, and where 
$G^{(\rho)}_{TS}(x)$ is equal to the previous modulo the
symbolic replacement $\rho\leftrightarrow\sigma$.
Here, $\theta_{\sigma,\rho} = 1/4$ and
$a/\xi_{\sigma,\rho} = \Delta_{\sigma,\rho}/v_{\sigma,\rho}$.
Hence, we arrive at the result
$\langle S_{x,1}^+ S_{0,1}^-\rangle \sim  e^{i 2k_F x}  (\alpha/x)^{1/2}
e^{-x/\xi_{xy}}$,  where
$\xi_{xy}/a = (\Delta_{\rho}/v_{\rho} + \Delta_{\sigma}/v_{\sigma})^{-1}$
is the (finite) $XY$ correlation length
that signals short-range transverse spin correlations.
Lorentz invariance in the dynamical
correlator $G^{(\rho,\sigma)}_{TS}(x,t)$ for
pseudo-triplet superconductivity
then implies that the dynamical  transverse spin
correlator (13) exhibits a spin gap
$\Delta_{xy} = \Delta_{\rho} + \Delta_{\sigma}$.
Similar calculations reveal that the longitudinal spin correlator has
the asymptotic form
$\langle S_{x,1}^z S_{0,1}^z\rangle = {\rm cos}(2k_F x) 
G_{SDW}^{(\rho)}(x) G_{SDW}^{(\sigma)}(x)/(4\pi\alpha)^2$, where
$G_{SDW}^{(\sigma)}(x)\rightarrow 1$ is the autocorrelator for
pseudo SDW order,$^7$ and where $G_{SDW}^{(\rho)}(x)$
is obtained again through a trivial symbolic replacement.
We therefore find that such Ising-coupled antiferromagnetic chains display
{\it strict} long-range order of the Ising type, like that
displayed  by an isolated Heisenberg chain in the presence of Ising
anisotropy.$^{4-6}$  
(See Table I for a listing of the relevant static correlation
exponents).
In conclusion, the
system is in a pinned SDW state that is commensurate with the lattice,
and that necessarily exhibits a spin gap.

The above results imply that this SDW state  is incommensurate$^9$ 
($k_F\neq \pi/2a$) for
large enough Zeeman energy splittings$^{10}$ $|h_z|\gg\Delta_{\rho}$,
in which case umklapp processes become irrelevant.
Then by Eq. (7), the pseudo-charge sector is in a pure
Luttinger liquid state, $\Delta_{\rho} \rightarrow 0$.
Also, since the bosonic field 
$\phi_{\rho} = 2^{-1/2}(\phi_1 + \phi_2)$ now 
represents a Goldstone
mode, where 
$\phi_l(x) = \phi_L (x,l) + \phi_R (x,l)$,
we have that the SDW can freely {\it slide} along the
chain direction.  The list of physical properties that we
outlined above must then be revised as follows.
Since the pseudo-charge system is now compressible, the 
longitudinal paramagnetic susceptibility no longer
shows a spin gap.  In particular, we have that 
$\chi_z = 
({\pi\over 2} J_{xy} + J_z^{\parallel} + {1\over 2} J_z^{\perp})^{-1}$ 
at zero temperature.  Likewise, the low-temperature specific heat
is now dominated by the pseudo-charge component
$C_{\rho}\propto T$.  Last, the sliding  Goldstone mode
results in only {\it algebraic} longitudinal
spin correlations
$\langle S_{x,1}^z S_{0,1}^z\rangle \sim {\rm cos}(2k_F x)
(\alpha/x)^{K_{\rho}}$, with exponent
$K_{\rho} = (2\pi v_{\rho} - g_{\rho})^{1/2}/(2\pi v_{\rho} + g_{\rho})^{1/2}$.
Note that here $S_z$ refers to the deviation of the magnetization
with respect to it's average value,$^{9,10}$
$\chi_z h_z$.
And although the transverse spin correlations remain finite in  the
the present incommensurate case, the $XY$ spin-correlation length,
$\xi_{xy}/a = v_{\sigma}/\Delta_{\sigma}$, is now
larger. 

We shall now treat the general case of $N\geq 2$ chains.  The previous
results, (9) and (10), obtained for the case of two Ising-coupled
chains indicate that the inter-chain forward scattering process (6)
can be neglected in the limit of weak coupling,
$|J_z^{\perp}|\ll J_{xy}^{\parallel}, |J_z^{\parallel}|$.  This
shall be  assumed throughout in the present discussion.  
To begin with, we shall also neglect all umklapp processes.  The Luttinger
model for the Ising-coupled antiferromagnetic chains in such
case reduces
to the sum of Eqs. (3) and (4), which describes a generalized
backscattering model.$^{14}$  A mean-field analysis$^{12}$ of this model
for large $N$
finds that long-range order of the charge-density  wave type,
$\langle L^{\dag}_l(x) R_l(x)\rangle\propto 
{\rm exp}[i(4\pi)^{1/2}\phi_l(x)]$,
is stable for $|J_z^{\perp}| > -2 J_z^{\parallel}$.
Notice that this agrees with the exact results for two
chains (see Fig. 1).
In addition, the application of the  abelian
bosonization technique,  (11) and (12), 
yields the action
$$S_{\rm LD} = i\int dx_0 dx_1
\Biggl\{\sum_{l=1}^N {1\over 2} (\partial_{\mu}\phi_l^{\prime})^2
-\xi_0^{-2}\sum_{l=1}^{N-1} {\rm cos} [(4\pi K)^{1/2} 
(\phi_{l+1}^{\prime} -\phi_l^{\prime})]
\Biggr\} \eqno (14)$$
related to the Lawrence-Doniach model of  layered superconductivity,$^{11}$ 
where $\phi_l^{\prime}(x,t)$ represents the
time evolution of the  bosonic field operator,
$\phi_l(x)$, after renormalization.$^6$
Here, $K = e^{2\psi}$  is one-half the exponent for
Ising order, with the angle $\psi$ set by the condition
${\rm tanh}\, 2\psi = - 2 J_z^{\parallel}/\pi J_{xy}^{\parallel}$,
while $x_{\mu} = (i  v_F^{\prime}  t, x)$ is a two-vector, 
with  a renormalized Fermi velocity
$v_F^{\prime} = J_{xy}^{\parallel}a\, {\rm sech}\, 2\psi$.
Also, $\xi_0\propto (J_z^{\perp})^{-1/2}$ denotes the
$XY$ spin-correlation length.
An analysis based on the equivalence of (14) to a layered Coulomb
gas$^{11}$  indicates that this term is relevant in the renormalization
group sense for $|J_z^{\perp}| > -4J_z^{\parallel}$.
Notice that this condition 
is roughly consistent with the phase diagram 
corresponding  to two chains (see Fig. 1),
as well as with
the mean-field result just cited.  Also, it is
clear from (14) that collective sliding
along the chain direction,
$\phi_l(x)\rightarrow\phi_l(x) + \delta\phi_{\rho}$, 
represents a Goldstone mode.
All of    these pieces of information when
put together then
strongly suggest that the groundstate of weakly Ising-coupled
chains corresponds to an unpinned
SDW state if umklapp processes are absent.

Suppose now that we include the umklapp processes 
present (at half-filling) in zero field.  The addition of the inter-chain
process (5) as well as
the intra-chain one$^{4-6}$
then leads to the  bosonic action
$$\eqalignno{
S = i \int d^2x \Biggl(\sum_{l=1}^N \Biggl\{ &
{1\over 2} (\partial_{\mu}\phi_l^{\prime})^2
-\xi_2^{-2} {\rm cos}[2(4\pi K)^{1/2} \phi_{l}^{\prime}]\Biggr\}\cr
& -\sum_{l=1}^{N-1} 2 \xi_0^{-2}
\,{\rm cos} [(4\pi K)^{1/2} \phi_{l+1}^{\prime}]
{\rm cos} [(4\pi K)^{1/2} \phi_{l}^{\prime}]
\Biggr),& (15)\cr
}$$
where $\xi_2$ represents the $XY$ spin-correlation
length of an isolated $XXZ$ chain.
If we then fix a particular chain $l$ and integrate
out the neighboring fields,$^{17}$ we obtain the effective 
action
$$\bar S_l = i \int d^2x 
\Bigl\{{1\over 2}
(\partial_{\mu}\phi_l^{\prime})^2
-\xi_{1}^{-2} {\rm cos}[(4\pi K)^{1/2} \phi_{l}^{\prime}]
-\xi_2^{-2} {\rm cos}[2(4\pi K)^{1/2} \phi_{l}^{\prime}]
\Bigr\},\eqno (16)$$
which is a sine-Gordon model$^{18}$ with a first harmonic
due to the intra-chain umklapp process.
Note that
the coefficient  $\xi_{1}^{-2}$
of the base term is not free, it being proportional to
the appropriate average
$\langle
{\rm cos} [(4\pi K)^{1/2} \phi_{l-1}^{\prime}]\rangle
+\langle 
{\rm cos} [(4\pi K)^{1/2} \phi_{l+1}^{\prime}]\rangle$
that results from the integration.
Within the equivalent Coulomb gas description, this 
inter-chain term  
corresponds to unit 
charges, while the  former 
intra-chain term proportional to $\xi_{2}^{-2}$  
corresponds to {\it double} charges.  Since double
charges in general dissociate into
unit charges if the system is in the (spin-gap) plasma phase,$^{19}$
we conclude that the first harmonic term is irrelevant
in the presence of the base term.  In the limit of
weak coupling, $J_z^{\perp}\rightarrow 0$,
this means that a spin-gap ($\xi_{1} < \infty$) opens for 
$J_z^{\parallel} > -(3\pi/10) J_{xy}^{\parallel}$, and that
bound states appear for $J_z^{\parallel} > 0$.$^{18}$
In contrast, a spin gap will  open only in the presence of Ising anisotropy,
$J_z^{\parallel} >  J_{xy}^{\parallel}$,
for the case of an isolated $XXZ$ chain ($J_z^{\perp} =  0$).$^6$
On the basis of what is known for a single $XXZ$ chain,$^{4,5}$
we conclude that Ising coupled spin-1/2 $XXZ$ chains with all-antiferromagnetic
couplings are generally in an Ising-antiferromagnetic state.
Notice that this claim relies on the validity of the reduction (16)
to a single spin chain,
which clearly fails in the case of  an {\it odd} number
of Heisenberg-coupled chains.$^{17}$
Nevertheless, the fact that we obtain a 
sliding SDW  state in the
absence of umklapp processes, coupled with the
fact that umklapp processes only reinforce (commensurate) SDW
order, strongly indicates that such a reduction is indeed
correct.

To conclude, we find that anisotropy in the spin-coupling that
may exist in between antiferromagnetic chains 
can dramatically change the nature of the groundstate of
each chain.
In particular, the present Luttinger model based analysis
demonstrates that consecutive spin-1/2 antiferromagnetic chains
flow to the Ising antiferromagnetic fixed point in the presence
of {\it any} amount of Ising coupling in  between 
nearest-neighbor chains (see Fig. 1). 
This contrasts with the case of
a single chain, where the Ising antiferromagnetic state  exist only
in the presence of Ising {\it anisotropy}, 
$J_z^{\parallel} > J_{xy}^{\parallel}$.
It also contrasts 
with the situation where
all couplings are Heisenberg,  in which  case
no spin gap exists for an odd number of chains.
The latter  suggests that a transition between the $XY$ and the Ising
antiferromagnetic fixed points
occurs at some intermediate perpendicular anisotropy,
$J_{xy}^{\perp} \lsim J_z^{\perp}$, in such case.

It is a pleasure to thank D. Poilblanc,
A. Nersesyan, and J. Fernandez-Rossier for discussions.
This work was supported in part by
National Science Foundation grant DMR-9322427.

\vfill\eject
\centerline{\bf References}
\vskip 16 pt

\item {1.} E. Dagotto and T.M. Rice, Science {\bf 271}, 618 (1996),
and references therein.

\item {2.} S.P. Strong and A.J. Millis, Phys. Rev. B {\bf 50}, 9911 (1994);
D.G. Shelton, A.A. Nersesyan, and A.M. Tsvelik, Phys. Rev. B {\bf 53}, 8521
(1996).

\item {3.} A. Luther and I. Peschel, Phys. Rev. B {\bf 12}, 3908 (1975).

\item {4.} F.D.M. Haldane, Phys. Rev. Lett. {\bf 45}, 1358 (1980).

\item {5.} M.P.M. den Nijs, Phys. Rev. B {\bf 23}, 6111 (1981).

\item {6.} E. Fradkin, {\it Field Theories of Condensed Matter Systems}
(Addison-Wesley, Redwood City, 1991), chap. 4.

\item {7.} V.J. Emery, in {\it Highly Conducting One-dimensional Solids},
ed. by J.T. Devreese, R.P. Evrard and V.E. van Doren
(Plenum Press, New York, 1979).

\item {8.} An SDW groundstate 
was conjectured for spin-1/2 $XXZ$ chains coupled via
Ising interactions in three dimensions;
see  V.J. Emery, Phys. Rev. B {\bf 14}, 2989 (1976).

\item {9.} G. Japaridze and A. Nersesyan, Pis'ma Eksp. Teor.
Fiz. {\bf 27}, 334 (1978); 
V.L. Pokrovskii and A.L. Talanov, Zh. Eksp. Teor. Fiz. {\bf 78},
269 (1980) [Sov. Phys. JETP {\bf 51}, 134 (1980)].

\item {10.} R. Chitra and T. Giamarchi, Phys. Rev. B {\bf 55}, 5816 (1997).

\item {11.} J.P. Rodriguez, J. Phys. Cond. Matter {\bf 9}, 
5117 (1997)(cond-mat/9604182); 
Europhys. Lett. {\bf 39}, 195 (1997) 
(cond-mat/9606154).

\item {12.} J.P. Rodriguez, ICMM-CSIC report (1997).

\item {13.} The particular case of two Ising-coupled $XY$
chains has been studied by
L. Hubert and A. Caill\' e, Phys. Rev. B {\bf 43}, 13187
(1991), using different methods.

\item {14.} A. Luther and V.J. Emery, Phys. Rev. Lett. {\bf 33}, 589
(1974).

\item {15.}  P.A. Lee, Phys. Rev. Lett. {\bf 34}, 1247 (1975).

\item {16.} S.T. Chui and P.A. Lee, Phys. Rev. Lett. {\bf 35}, 315
(1975).

\item {17.} H.J. Schulz, Phys. Rev. Lett. {\bf 77}, 2790 (1996).

\item {18.} S. Coleman, Phys. Rev. D {\bf 11}, 2088 (1975).

\item {19.} A.M. Polyakov, Nucl. Phys. B{\bf 120}, 429 (1977);
{\it Gauge Fields and Strings} (Harwood, New York, 1987).

\vfill\eject
\centerline{\bf Figure Caption}
\vskip 20pt
\item {Fig. 1.}  Shown is the phase diagram for two spin-1/2
$XXZ$ chains coupled via a weak Ising interaction ($J_z^{\perp}$).
The Ising antiferromagnetic (AF) regions
($\Delta_{\rho,\sigma}\neq 0$)
 are characterized by strict
long-range spin order along each chain, while the $XY$ AF regions
($\Delta_{\rho,\sigma}=0$)
show dominant $XY$ spin correlations
$\langle S_{x,l}^+ S_{0,l}^-\rangle\propto 
e^{i2k_F x}(\alpha/x)^{(K_{\rho}^{-1}
+ K_{\sigma}^{-1})/4}$ along each chain $l$
[see Eq. (13)].
The sign of the intra-chain Ising coupling ($J_z^{\parallel}$)
switches between these two phases by passing through
an intermediate $XY$ dimer phase
($\Delta_{\rho} \neq 0, \Delta_{\sigma} =  0$)
 characterized by dominant 
inter-chain dimer correlations
$\langle S_{x,1}^+ S_{x,2}^- S_{0,2}^+ S_{0,1}^-\rangle
\propto (\alpha/x)^{K_{\sigma}^{-1}}$ for antiferromagnetic
coupling $J_z^{\perp} > 0$.  Quadrapolar $XY$ spin correlations
$\langle S_{x,1}^+ S_{x,2}^+ S_{0,2}^- S_{0,1}^-\rangle
\propto (\alpha/x)^{K_{\rho}^{-1}}$
dominate in the opposing region
($\Delta_{\rho} = 0, \Delta_{\sigma} \neq  0$)
 at ferromagnetic couplings.
Note that the superscript $^{(+)}$ indicates local ferromagnetic order
in between chains.

\bigskip\bigskip\bigskip\bigskip\bigskip

\vfill\eject
\pageno = 12

\noindent {TABLE I.}  Listed is the correlation exponent $\eta$
obtained via the bosonization technique$^3$ for various
order parameters, $O(x)$, in the spin ladder; i.e., 
$\langle O(x) O^{\dag}(0)\rangle\propto (\alpha/x)^{\eta}$.  
Antiferromagnetic Ising coupling in between chains is assumed
(see Fig. 1).
Note that the value $\eta = 0$ indicates strict long-range order, while
$\eta = \infty$ indicates short-range order.  Below, we have
$K_{\rho,\sigma} = (2\pi v_{\rho,\sigma} - g_{\rho,\sigma})^{1/2}
/(2\pi v_{\rho,\sigma} + g_{\rho,\sigma})^{1/2}$.

\bigskip\bigskip\bigskip

\vbox{\offinterlineskip
\hrule
\halign{&\vrule#&
  \strut\quad\hfil#\hfil\quad\cr
height2pt&\omit&&\omit&&\omit&&\omit&\cr
&Order Parameter\hfil&&$\eta$\quad($XY$ AF)\hfil&&$\eta$\quad($XY$ Dimer)\hfil
&&$\eta$\quad(Ising AF)&\cr
height2pt&\omit&&\omit&&\omit&&\omit&\cr 
\noalign{\hrule}
height2pt&\omit&&\omit&&\omit&&\omit&\cr 
&$S_{i,l}^y$
&&${1\over 4}(K_{\rho}^{-1} + K_{\sigma}^{-1})$&&$\infty$&&$\infty$&\cr
&$S_{i,l}^z$&&$K_{\rho} + K_{\sigma}$&&$K_{\sigma}$&&$0$&\cr
&$S_{i,1}^+ S_{i,2}^-$&&$K_{\sigma}^{-1}$&&$K_{\sigma}^{-1}$&&$\infty$&\cr 
&$S_{i,1}^+ S_{i,2}^+$&&$K_{\rho}^{-1}$&&$\infty$&&$\infty$&\cr  
height2pt&\omit&&\omit&&\omit&&\omit&\cr}
\hrule}

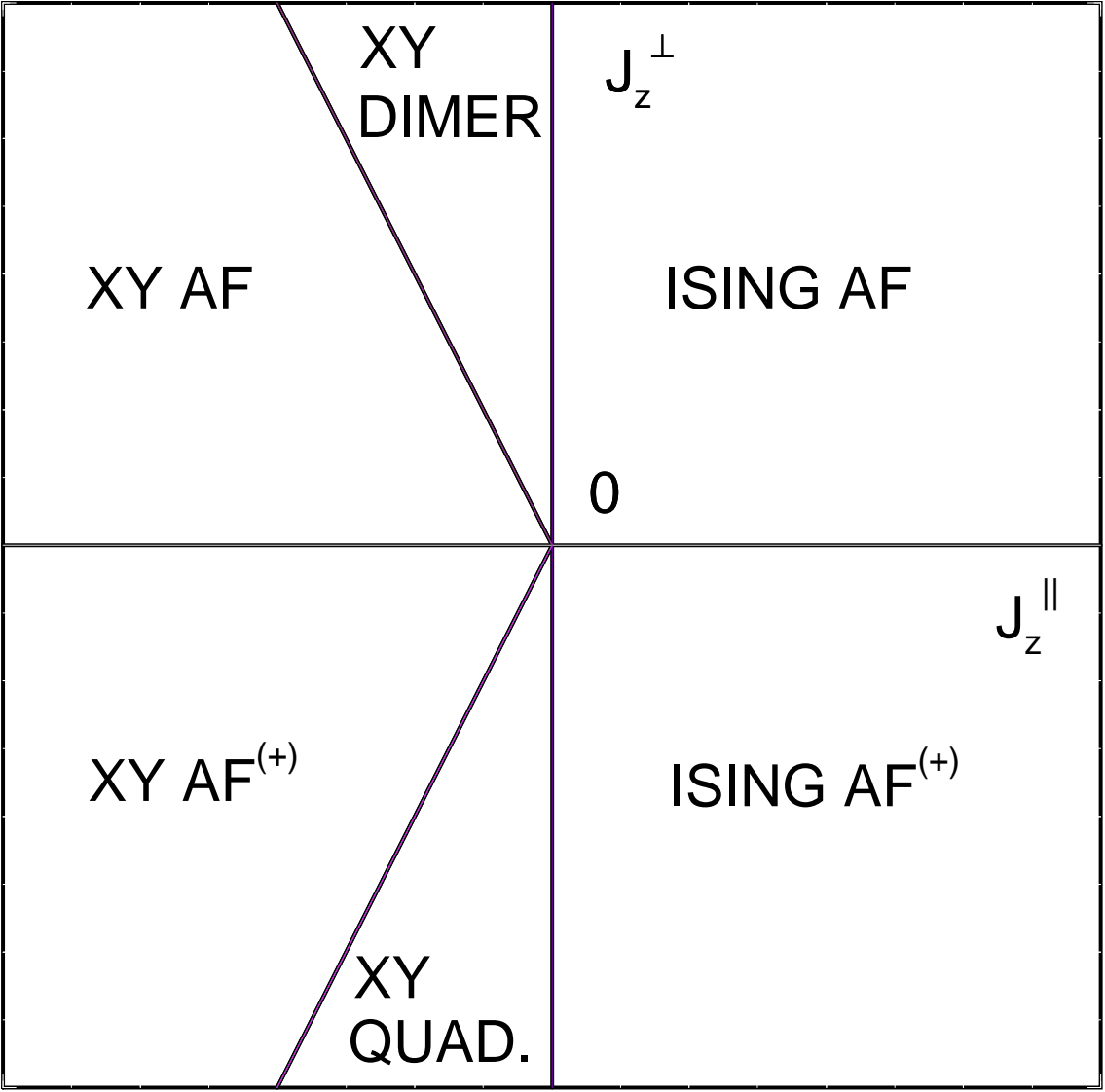
\end